\documentclass[12pt]{article}
\usepackage{graphicx}
\usepackage{amssymb}
\usepackage{epstopdf}
\usepackage{color}
\def\beq{\begin{equation}}
\def\eeq{\end{equation}}
\def\bea{\begin{eqnarray}}
\def\eea{\end{eqnarray}}

\def\bnotes{\noindent {\it Comments for Further Consideration in this Section}\begin{itemize}}
\def\enotes{\end{itemize}}

\def\msection#1{\section{\large #1}}

\begin{document}

%\title{\begin{flushright}
%{\small CERN-PH-TH-2011-138\\
%MCTP-2011-22}
%\end{flushright}

\title{\bf\large Effective Field Theories and \\ the Role of Consistency in Theory Choice}
%Inference to the Best Explanation}
\author{James D. Wells\\
{\it CERN Theoretical Physics (PH-TH), CH-1211 Geneva 23, Switzerland} \\
{\it Department of Physics, University of Michigan, Ann Arbor, MI 48109}}
\date{}

\maketitle

\begin{abstract}

Promoting a theory with a finite number of terms into an effective field theory with an infinite number of terms worsens simplicity, predictability, falsifiability, and other attributes often favored in theory choice.  However, the importance of these attributes pales in comparison with {consistency}, both {observational} and {mathematical consistency}, which propels the effective theory to be superior to its simpler truncated version of finite terms, whether that theory be renormalizable (e.g., Standard Model of particle physics) or nonrenormalizable (e.g., gravity).  Some implications for the Large Hadron Collider and beyond are discussed, including comments on how directly acknowledging the preeminence of consistency can affect future theory work.

\end{abstract}
                                           % Activate to display a given date or no date

\vspace{0.1in}

\begin{center}
{\it 
Lecture delivered at Wuppertal University for the physics and philosophy conference 
``The Epistemology of the Large Hadron Collider", 26-28 January 2012.}
\end{center}

\vfill\eject

\tableofcontents

\vfill\eject

%%%%%%%%%%%%%%%%%%%%%%%%%%%%%%%%%%%%%%%%%%%%
\msection{Introduction}

One of the most interesting questions in philosophy of science is how to determine the quality of a theory. Given the data, how can we infer a ``best explanation" for the data. This often goes by the name ``Inference to Best Explanation" (IBE)~\cite{Harman:1965,Lipton:1991,Clayton:1997}.  The wide variety of claims for important criteria are a measure of how difficult it is to come up with a clear and general algorithm for choosing between theories. Some claim even that it is intrinsically not possible to come up with a methodology of deciding~\cite{Lehrer:1974,Newton-Smith:1999}. Nevertheless the goals of IBE are worthy, and the payoff is high upon increased understanding, if for no other reason than the extraphilosophical importance of distributing grant money more fairly to researchers. Furthermore, whether objective criteria for IBE are possible, all practitioners of science have no choice but to engage in the ``infer" part even if they may never touch upon the ``best explanation" part of IBE. 
%So carry on we must.

The goal of this article is to survey theory choice criteria in the context of effective theories. It has been accepted by the physics communities that theories must be ``effectified", that is they must be augmented to include all possible interactions consistent with the stated symmetries to all orders. On the surface the resulting effective theories are in conflict with the rules of IBE, whether they be the murky rules that some physicists put forward when they talk  about theory choice, or the precisely stated rules developed by philosophers.  Upon closer inspection effective theories rise quickly to the top in theory choice when admitting to the primacy of {\it consistency} in theory choice. That is the claim, to be developed below. The reader should be warned that I will use the acronym IBE to mean any attitude, theory, system by which people decide that one theory is a better description of nature than another, or that a theory under consideration is a good theory at all.

\msection{The Standard Model's Triumphs and Woes}

This article is primarily written from the science perspective of elementary particle theory, with particular emphasis on the subfield ``beyond the Standard Model physics". In this subfield, the task is to look out over nature and ask what is not adequately described by the Standard Model (SM) of particle physics. The Standard Model has been with us for about 40 years. It consists of three families of up-type quarks ($u,c,t$), three families of down-type quarks ($d,s,b$), three families of leptons $(e,\mu,\tau)$ and three families of neutrinos $(\nu_e,\nu_\mu,\nu_\tau)$. These interact with each other according to gauge field theory interactions, mediated by the force carrier bosons of the photon, gluons and $W$ and $Z$ bosons. Every particle that has mass is said to achieve it by a condensing Higgs boson. For a more complete non-technical or technical description of the SM see refs.~\cite{Kane:1996} and~\cite{Griffiths:2008} respectively.

The SM is a renormalizable theory and can be fully described on one page using standard nomenclature of mathematics and relativistic quantum field theory.
Despite that simplicity, it can account for every measurement ever made at high-energy colliders. It is an enormous human achievement. So why are there researchers searching for theories ``beyond the SM"? There are many reasons, of which I will name a few: 
\begin{itemize}
\item  There are non-collider observations we still cannot explain such as galactic rotation measurements that imply the existence of dark matter, and  the preponderance of baryons over anti-baryons in the universe. 
\item The particle content and the three gauge forces  cry out for unification (e.g., grand unified theories)
\item There are many of parameters with large hierarchies that beg for explanation ($m_t/m_e>10^6$).
\item The SM Higgs boson appears unstable to quantum corrections and is thus unnatural.
\item Surely there is more than just the SM (e.g., SM is just copies of stuff in our bodies).
\item Embedding gravity into quantum mechanics is a severe challenge and should bring new implications to the particle physics world (e.g., string theory)
\end{itemize}
Thus, there are many opportunities to devise new theories that solve one or more of these problems. The theories are necessarily speculative upon birth. They are put to the test, and the simple fact is that at any given moment there are a multitude of theories that appear to be able to solve one or more of the issues.   We have many variants of supersymmetric theories, strongly coupled theories, extra dimensional theories, etc.\ that appear to be able to do the job and are not yet distinguishable by currently known data.  How does a scientist determine which is the best? The rules are not clear, of course, and we shall first ask how do scientists make theory choices, and when does IBE enter their calculus.

%%%%%%%%%%%%%%%%%%%%%%%%%%%%%%
\msection{Theory Choice among Practitioners}

Typically a particle physicist will look at the SM problems listed above and set out to construct a new theory that explains one or more of them. The particle researcher often stumbles into a theory choice of what to work on not based on IBE but rather DBO (deduction of best opportunities). The opportunities that arise may include matching yourself with the best PhD advisor who is working on theory X, researching a fashionable topic to get a good job, supporting a clever theory that the researcher devised that might not have high probability of being correct but has highest probability of enormous personal pay-off, etc. The last reason then circles back on the first reason as advisors ask their students to work on theories that they themselves devised. Furthermore, the subtleties of elementary particle physics and beyond the SM theories are such that it could take new practitioners years before they feel confident that they could make a reliable IBE estimate even if the criteria for such were clear to them. Thus, IBE considerations are often not the dominant force for their theory choice (i.e., what to work on) in a practicing scientist's career. 

IBE issues do arise when there is competition among researchers for journal space, research funds, and conference time slots. IBE-like arguments ensue. Words used to describe the evaluation of theories are familiar to philosophers: simplicity, economy, calculability, compatibility with data, testability, falsifiability, naturalness, finetuning, predictivity, unification, no ad hoc assumptions, etc. Researchers become attorneys for their theories and weight the various IBE criteria which most favorably supports the direction of their research lines. That is why experimentalists and phenomenologists emphasize ``falsifiability" and ``observational consistency" much more than string theorist, who emphasize ``unification", ``completeness" and ``mathematical consistency".

It is often said at the end of arguments between theorists about their pet theories that  ``experiment will decide". However, as experiments become larger and costs grow steeply, the time frame may extend well past decades to even centuries. It took more than 25 years for CERN to conceive and build the LHC, for example. There is no guarantee that any timeline convenient to a human  is relevant to future experimental construction. However, what is relevant to human time scales is deciding what are ``better" or ``best theories", since we should use that to allocate resources of time, money, etc. Working toward perfecting IBE criteria, no matter how controversial they are, is clearly warranted.  

The recent universal acclaim of effective theories gives us an opportunity to apply IBE thinking to a case that is not controversial as a means to better understand the weight that should be given to various elements of IBE.  In the next section, I will describe how the effective SM is different from the SM, and then we shall survey their IBE qualities, with an eye toward gaining insight along the way.

%%%%%%%%%%%%%%%%%%%
\msection{The Standard Model vs.\ the Effective Standard Model}

The SM has a finite number of operators of dimension four or less. The effective SM (ESM) is the SM but with all possible higher dimensional operators present consistent with the sacrosanct symmetries of the SM: Lorentz symmetry and $SU(3)_c\times SU(2)_L\times U(1)_Y$ gauge symmetries.  Thus we can relate the lagrangians of the two by the equation
\beq
{\cal L}_{ESM}={\cal L}_{SM}+\sum_{n,i}\eta_{i}\frac{{\cal O}^{(4+n)}_i}{\Lambda^n},
\eeq
where ${\cal O}^{(4+n)}_i$ is the collection of all operators of higher dimension $4+n$ that respect the symmetries of the SM and have unknown couplings $\eta_{i}/\Lambda^n$ in front, where $\Lambda$ is the cutoff scale and $\eta_i$ are numerical pre-factors that can be different for each operator.

The SM matches all observed high-energy collider data to excellent compatibility. There can be no additional operators that would improve the fit by a meaningful amount. Furthermore, if any of the higher-dimensional operators of ${\cal L}_{ESM}$ become worrisome with respect to the data, we need merely tune down the strength of the interaction by making its associated $\eta_{n,i}$ coupling smaller, to escape the problem.  

Which theory is better, the ESM or SM, given that they both can be made equally compatible with the data?  To answer this question, let us first apply some of the IBE thinking common in the particle physics community. Our example source for a typical particle physicist approach to these issues will be the essay written by Nobel Laureate Burton Richter~\cite{Richter:2006}. We shall also attempt to answer the question using the criteria of the philosopher Paul Thagard~\cite{Thagard:1978}, whose paper is still considered one of the key early expositions on theory choice criteria for IBE.

%%%%%%%%%%%%%%%%%%%%
\msection{Richter's IBE Criteria}

There are not many official forums through which practicing particle physicists are encouraged to divulge their IBE criteria. But one forum where it regularly happens, both in essays and in letters to the editor, is in more informal publications such as professional society monthly magazines and newsletters. One of the most talked about articles of this kind in recent years was written by Burton Richter~\cite{Richter:2006}. Richter and Sam Ting won the Nobel prize of physics in 1976 for finding the $J/\psi$ particle, which was a key discovery in establishing the SM. 

Richter  has been horrified by what he views are ``major problems in the philosophy behind theory" research. He says,
\begin{quote} 
Simply put, most of what currently passes as the most advanced theory looks to be more theological speculation, the development of models with no testable consequences, than it is the development of practical knowledge, the development of models with testable and falsifiable consequences (Karl Popper's definition of science).
\end{quote}
Richter goes on to say that more weight should be put on
\begin{enumerate}
\item theories that have testable and {\it falsifiable} consequencies, and
\item  theories that {\it simplify} rather than increase complication.
\end{enumerate}
Incidentally, he also discusses two anti-criteria that  should not be used, which are the anthropic principle and naturalness. Let us not discuss these anti-criteria, but rather judge the SM vs.\ ESM based on what Richter would have us do, on falsifiability and simplicity.

Regarding falsifiability and testable consequences, an argument can be made that the SM wins. The ESM has an infinite number of operators with coefficients to be pinned down by data later, and as such can accommodate more experimental outcomes than the SM.  After all, the ESM reduces to SM when $\Lambda^n\to \infty$. Thus, the SM is much more testable and falsifiable than ESM.

Although not central to the subsequent discussion, I would like to remark that {\it falsifiability} has never struck me as strong argument for theory deciding for two reasons. Skepticism toward falsifiability has long been held in the philosophy community, but let me give what I think are two strong reasons to worry about its applicability. Let's take an example of an unfalsifiable theory: Theory X says that obvious fact Y is true (e.g., emeralds are green, or something trivially true like that), {\it and} that angels live in another universe.  We can use this silly theory to illustrate why falsifiability is not a very solid criteria. 

First, the modularity of the theory can be under dispute, such as the more testable first statement versus the second statement. Second, if things change dramatically such that what was true yesterday is not true tomorrow (tomorrow Y is false), then the theory is trivially invalidated. Does that make it falsifiable? In that case, all theories are falsifiable by scattering the word ``always" through-out its description.  And third, it is never clear if {\it falsifiability} must be applicable in principle or in practice. In principle perhaps everything is testable and {\it falsifiable} (e.g., many versions of string theory -- just run a collider at $10^{19}\, {\rm GeV}$), whereas in practice good theories might not be (perhaps: string theory, high scale warped extra dimensions, etc.)\ because of lack of technology, time, money or manpower to test it.  

In short, you can like it, you can hope for it, you can wish for it, you can say it would make our lives as scientists much easier if so, but it would presumptuous of us to say that Nature cares one whit if we can falsify a true statement.  Nevertheless, we shall take it seriously because we are investigating somebody else's criteria, which happen to be shared by many others. And as we have noted above, the testable and falsifiability criteria favors SM over ESM.

Regarding {\it simplicity}, the SM has a finite number of terms with a finite number of coefficients, whereas the ESM has an infinite number of terms with an infinite number of coefficients. No contest, SM wins in the simplicity category.

According to Richter's two key criteria, falsifiability and simplicity, the SM is the  winner, and we infer it to be the best theory.

%%%%%%%%%%%%%%%%%%%%%%%%%%%%
\msection{Thagard's IBE Criteria\label{sec:thagard}}

The philosophy literature is vast on the subject. Surveying it with sweeping scope would not be enlightening and picking just one approach to compare leaves one wanting. Nevertheless, I will do the latter, choosing a classic paper on the subject by Paul Thagard~\cite{Thagard:1978}.  

Thagard's theory choice criteria are 
\begin{enumerate}
\item {\it Consilience}: The measure of how many facts the theory explains; furthermore, ``a consilient theory unifies and systematizes."
\item {\it Simplicity}: The quality of having the fewest ``auxiliary hypothesis," fewest ad hoc additions, and most ontological economy. 
\item {\it Analogy}: Shared characteristics between two theories, leads to one theory admitting a new characteristic if the new characteristic is part of the other theory and explains the shared characteristics there.
\end{enumerate}

Regarding {\it consilience}, it is a draw between SM and ESM. The facts are equally compatible in the two theories, and there is no relevant advantage in either in the realm of unification and systematizing.

Regarding {\it simplicity}, although it is not exactly the kind of simplicity that Richter was talking about, the SM clearly is superior to the ESM in this category. The new operators of the ESM simply add more.

Regarding {\it analogy}, it is my view that the ESM wins out over the SM.  I will explain why twice. First, I will explain it here strictly in the language of Thagard's analogy propositions. I will explain it a second time later heuristically using particle physics language from Steven Weinberg.

As Thagard explains, by {\it analogy} he does not mean the standard syllogism\\
\indent\indent \indent\indent $A$ is $P,Q,R,S$ \\
\indent\indent \indent\indent $B$ is $P,Q,R$ \\
\indent\indent \indent\indent Thus, $B$ is $S$.\\
No, he means something more causally connected:\\
\indent\indent \indent\indent $A$ is $P,Q,R,S$ \\
\indent\indent \indent\indent $B$ is $P,Q,R$ \\
\indent\indent \indent\indent If $S$ explains $P,Q,R$ in $A$, then $B$ is $S$.\\

I will follow this analogy criteria by first definining \\
\indent\indent \indent$A\, =$  chiral lagrangian of pion scattering \\
\indent\indent \indent$B\, =$  Effective Standard Model (ESM). \\
The chiral lagrangian of pion scattering is
\bea
{\cal L}_{\pi}=\frac{v^2}{4}Tr(\partial_\mu U\partial^\mu U^\dagger)+\frac{v^2}{\Lambda^2}
\left[ Tr(\partial_\mu U\partial^\mu U^\dagger)\right]^2+\cdots
\eea
where
\bea
U\equiv \exp( i \vec \tau\cdot \vec \pi/v^2).
\eea
This lagrangian has an infinite number of terms respecting the underlying $SU(2)$ custodial symmetries. The pions are the $\vec \pi$ fields in $U$, and $v$ is the vacuum expectation breaking of the custodial symmetry $SU(2)_L\times SU(2)_R$ to its $SU(2)_V$ vector subgroup. All the interactions of the pions are contained within these terms. As the energy increases the higher order terms in the lagrangian become more important, and the data can be accommodated. This theory was very useful.  It was determined that all the higher order corrections needed to be there, although a deep appreciation of why was not to come until Ken Wilson's renormalization breakthroughs years later.

Now, the shared properties $P,Q,R$ of the chiral lagrangian theory of pion scattering and the ESM are 
\begin{quote} 
$P,Q,R\, =$ quantum field theory, perturbative expansion theory, all lowest dimensionality terms allowed by symmetries of the theory are present, finite number of terms relevant in deep infrared, etc.
\end{quote}
and the new characteristic $S$ in theory $A$ is 
\begin{quote}
$S\, =$ all operators consistent with the symmetries are present
\end{quote}
$S$ explains $P,Q,R$ because relevant terms are a subset of ``all operators".  

Thus, by Thagard's {\it analogy} we would say that the SM should be augmented by all possible terms consistent with its symmetries $\longrightarrow$ ESM. The argument is further strengthened later when we catch Weinberg directly using the language of analogy to support the generalization of effective field theory techniques to the SM.

The result of our analysis based on Thagard's IBE criteria is SM $+1$, ESM $+1$, and Draw $+1$. No clear resolution to be found here, although compared to Richter's criteria there is more support of the ESM in Thagard's approach.

%%%%%%%%%%%%%%%%%%
\msection{Non-negotiable attributes of a Best Explanation}

What is lacking in our discussion of IBE criteria is a rank ordering of attributes.  We must first ask ourselves what is non-negotiable. {\it Falsifiability} is clearly something that can be haggled over. {\it Simplicity}  is subject to definitional uncertainty, and furthermore has no universally accepted claim to preeminence. {\it Naturalness}, {\it calculability}, {\it unifying ability}, {\it predictivity}, etc.\ are also subject to preeminence doubts.

What is non-negotiable is {\it consistency}.  A theory shown definitively to be inconsistent does not live another day. It might have its utility, such as Newton's theory of gravity for crude approximate calculations, but nobody would ever say it is a better theory than Einstein's theory of General Relativity\footnote{The word ``better" in this context can induce apoplectic shocks in pedants. To avoid that, by ``better" I wish to say that it is closer to the true, underlying theory, whatever that may mean or be. I do not wish it to mean ``better to calculate a hammer fall on the moon in under three lines for primary school children", or any other similar appeal to convenience or simplicity.}.

{\it Consistency} has two key parts to it. The first is that what can and has been computed must be consistent with all known observational facts. As Murray Gell-Mann said about his early graduate student years, ``Suddenly, I understood the main function of the theoretician: not to impress the professors in the front row but to agree with observation~\cite{Gell-Mann:1994}."
Experimentalists of course would not disagree with this non-negotiable requirement of {\it observational consistency}. If you cannot match the data what are you doing, they would say? 

However, theorists have a more nuanced approach to establishing {\it observational consistency}. They often do not spend the time to investigate all the consequences of their theories. Others do not want to ``mop up" someone else's theory, so they are not going to investigate it either. We often get into a situation of a new theory being proposed that solves one problem, but looks like it might create dozens of other incompatibilities with the data but nobody wants to be bothered to compute it. Furthermore, the implications might be extremely difficult to compute. 

Sometimes there must be suspended judgment in the competition between excellent theories and observational consequences. Lord Kelvin claimed Darwin's evolution ideas could not be right because the sun could not burn long enough to enable long-term evolution over millions of years that Darwin knew was needed. Darwin rightly ignored such arguments, deciding to stay on the side of geologists who said the earth appeared to be millions of years old~\cite{Gavin}.  Of course we know now that Kelvin made a bad inference because he did not know about the fusion source of burning within the sun that could sustain its heat output for billions of years.

A second part to {\it consistency} is {\it mathematical consistency}. There are numerous examples in the literature of subtle mathematical consistency issues that need to be understood in a theory. Massive gauge theories looked inconsistent for years until the Higgs mechanism was understood. Some gauge theories you can dream up are ``anomalous" and inconsistent. Some forms of string theory are inconsistent unless there are extra spatial dimensions. Extra time dimensions appear to violate causality, even when one tries to demand it from the outset, thereby rendering the theory inconsistent. Theories with ghosts, which may not be obvious upon first inspection, give negative probabilities of scattering. 

{\it Mathematical consistency} is subtle and hard at times, and like {\it observational consistency} there is no theorem that says that it can be established to comfortable levels by theorists on time scales convenient to humans. Sometimes the inconsistency is too subtle for the scientists to see right off. Other times the calculability of the mathematical consistency question is too difficult to give definitive answer and it is a ``coin flip" whether the theory is ultimately consistent or not. For example, pseudomoduli potentials that could cause a runaway problem are incalculable in some interesting dynamically broken supersymmetric theories~\cite{Intriligator:2008fe}. 

It is not controversial that {\it observational consistency} and {\it mathematical consistency} are non-negotiable; however, the due diligence given to them in theory choice is often lacking. 
The establishment of {\it observational consistency} or {\it mathematical consistency} can remain in an embryonic state for years while research dollars flow and other IBE criteria become more motivational factors in research and inquiry, and the {\it consistency} issues become taken for granted.

This is one of the themes of Gerard `t Hooft's essay ``Can there be physicst without experiments?"~\cite{562147}. He reminds the reader that some of the grandest theories are investigations of the nature of spacetime at the Planck scale, which is many orders of magnitude beyond where we currently have direct experimental probes.  If this is to continue as a physics enterprise it ``may imply that we should insist on much higher demands of logical and mathematical rigour than before."  Despite the weakness of verb tense employed, it is an incontestable point. It is in these Planckian theories, such as string theory and loop quantum gravity, where the lack of {\it consistency} rigor is so plainly unacceptable. However, the cancer of lax attention to consistency can spread fast in an environment where theories and theorists are f\^eted before vetted.

%%%%%%%%%%%%%%%%%%%%%%%%%%%%%%%%%
\msection{Effective field theories and consistency}

Let us begin with the claim at the heart of our discussion. The claim behind the ascendancy of effective theories is that unless there is good and explicit reason otherwise, consistency requires that a theory have all possible interactions consistent with its symmetries at every order.  

The claim has its origins in the work of Wilson, whose original review article with Kogut~\cite{Wilson:81239} is a classic. There are many modern reviews of effective theories that make or assume the above claim~\cite{Polchinski:1992ed,Cohen:1993,Rothstein:2003}.  Weinberg's recent historical perspective~\cite{Weinberg:2009bg} gives an excellent summary of what was learned:
\begin{quote}
I was struck [at Erice school in 1976] by Kenneth Wilson's device of ``integrating out" short-distance degrees of freedom by introducing a variable ultraviolet cutoff, with the bare couplings given a cutoff dependence that guaranteed that physical quantities are cutoff independent. Even if the underlying theory is renormalizable, once a finite cutoff is introduced it becomes necessary to introduce every possible interaction, renormalizable or not, to keep physics strictly cutoff independent.... Indeed, I realized that even without a cutoff, as long as every term allowed by symmetries is included in the Lagrangian, there will always be counterterm available to absorb every possible ultraviolet divergence....
\end{quote}
Therefore, consistency of the theory -- the absorption of ultraviolet divergences, the maintaining of independence of arbitrary ultraviolet scale cutoff, etc. -- requires the introduction of all possible terms allowed by the symmetries. 

The issue of consistency then becomes front and center, and the issues of simplicity and testability fade in importance.  From our discussion above we know that without this important issue of consistency, the effective SM may not win in a theory choice competition compared to the SM with just its renormalizable operators, since it worsens the otherwise positive features of simplicity and testability. Therefore, the establishment of rigorous consistency requirements on the theory were crucial in the decision.

%%%%%%%%%%%%%%%%%%%%%%%%%%%%%%%%%%%%%
\msection{Relation to Thagard's analogy criterion}

I would like to take a quick aside and show that physicists do reason in real-life, complex theory circumstances through the {\it analogy} criterion of Thagard. Indeed, it is a separate argument for the general applicability of effective theories.

In the same historical review article~\cite{Weinberg:2009bg} quoted above, Weinberg shows that because effective field theory ideas were necessary in chiral dynamics (low-energy pion scattering), the concept should also apply to the SM.  Here is a relevant quote:
\begin{quote}
Perhaps the most important lesson from chiral dynamics was that we should keep an open mind about renormalizability. The renormalizable Standard Model of elementary particles may itself be just the first term in an effective field theory that contains every possible interaction allowed by Lorentz invariance and the $SU(3)\times SU(2)\times U(1)$ gauge symmetry, only with the non-renormalizalbe terms suppressed by negative powers of some very large mass $M$, just as the terms in chiral dymamics with more derivatives ... are suppressed by negative powers of $2\pi F_\pi\simeq m_N$.
\end{quote}
One should note the usage of analogy language: ``most important lesson from chiral dynamics" and ``just as in the terms in chiral dynamics". Thus, the syllogistic representations given in sec.~\ref{sec:thagard} are shown to apply and be part of theory choice for particle physicists.

%%%%%%%%%%%%%%%%%%%%%%%%%%%%%%%%%
\msection{Summary: the preeminence of consistency}

I will conclude by stating my two central points that generalize the discussion we have had above in comparing the effective SM with the SM.

My first point is that the conditions of theory choice should be ordered.  
Frequently we see the listing of criteria for theory choice given in a flat manner, where one is not given precedence over the other a priori. We see consilience, simplicity, falsifiability, naturalness, consistency, economy, all together in an unordered list of factors when judging a theory. 
However, {\it consistency} must take precedence over any other factors. Observational consistency is obviously central to everyone, most especially our experimental colleagues, when judging the relevance of theory for describing nature.  Despite some subtleties that can be present with regards to observational consistency\footnote{There can be circumstances where a theory is observationally consistent in a vast number of observables, but in a few it does not get right, yet no other decent theory is around to replace it. In other words, observational consistency is still the top criterion, but the best theory may not be 100\% consistent.} it is a criterion that all would say is at the top of the list.

Mathematical consistency, on the other hand, is not as fully appreciated. In Richter's essay excoriating theorists, he did not appear to recognize or acknowledge the central role that mathematical consistency plays in developing and vetting theories.  Mathematical consistency has a preeminent role right up there with observational consistency, and can be just as subtle, time-consuming and difficult to establish.  We have seen that in the case of effective theories it trumps other theory choice considerations such as simpleness, predictivity, testability, etc.

My second point builds on the first. Since consistency is preeminent, it must have highest priority of establishment compared to other conditions. Deep, thoughtful reflection and work to establish the underlying self-consistency of a theory takes precedence over finding ways to make it more natural or to have less parameters (i.e., simple). Highest priority must equally go into understanding all of its observational implications. A theory should not be able to get away with being fuzzy on either of these two counts, before the higher order issues of simplicity and naturalness and economy take center stage. That this effort might take considerable time and effort should not be correlated with a theory's value, just as it is not a theory's fault if it takes humans decades to build a collider to sufficiently high energy and luminosity to test it.

Additionally, dedicated effort on mathematical consistency of the theory, or class of theories, can have enormous payoffs in helping us understand and interpret the implications of various theory proposals and data in broad terms.  An excellent example of that in recent years is by Adams et al.~\cite{hep-th/0602178}, who showed that some theories in the infrared with a cutoff cannot be self-consistently embedded in an ultraviolet complete theory without violating standard assumptions regarding superluminality or causality.  The temptation can be high to start manipulating uninteresting theories into simpler and more beautiful versions before due diligence is applied to determine if they are sick at their cores. This should not be rewarded.

%%%%%%%%%%%%%%%%%%%%%%%%%%%%%%%%%%
\msection{Implications for the LHC and beyond}

Finally, I would like to make a comment about the implications of this discussion for the LHC and other colliders that may come in the future. First, it is obvious that we must be prepared for and search for higher-dimensional operators in the effective SM that go beyond the %relevant and marginal 
operators of the SM.  This is indeed happening at the LHC, and first indications of new physics may very well come from small perturbations in SM observables due to the subtle effects of these suppressed operators.

However, there is broader point to be made regarding implications for colliders.  In the years since the charm quark was discovered in the mid 1970's there has been tremendous progress experimentally and important new discoveries, including the recent discovery of a Higgs boson-like state~\cite{HiggsDiscovery}, but no dramatic new discovery that can put us on a straight and narrow path beyond the SM. That may change soon at the LHC. Nevertheless, it is expensive in time and money to build higher energy colliders, our main reliable transporter into the high energy frontier. This limits the prospects for fast experimental progress. 

In the meantime though, hundreds of theories have been born and have died. Some have died due to incompatibility of new data (e.g., simplistic technicolor theories, or simpleminded no-scale supersymmetry theories), but others have died under their own self-consistency problems (e.g., some extra-dimensional models, some string phenomenology models, etc.). In both cases, it was care in establishing consistency with past data and mathematical rigor that have doomed them. In that sense, progress is made. Models come to the fore and fall under the spotlight or survive. When attempting to really explain everything, the consistency issues are stretched to the maximum. For example, it is not fully appreciated in the supersymmetry community that it may even be difficult to find a ``natural" supersymmetric model that has a high enough reheat temperature to enable baryogenesis without causing problems elsewhere~\cite{WellsPapers}. There are many examples of ideas falling apart when they are pushed very hard to stand up to the full body of evidence of what we already know.

Relatively speaking, theoretical research is inexpensive. It is natural that a shift develop in fundamental science. The code of values in theoretical research will likely alter in time, as experimental input slows. Ideas will be pursued more rigorously and analysed crtically. Great ideas will always be welcome. However, soft model building tweaks for simplicity and naturalness will become less valuable than rigorous tests of mathematical consistency. Distant future experimental implications identified for theories not fully vetted will become less valuable than rigorous computations of observational consistency across the board of all currently known data. One can hope that unsparing devotion to full consistency, both observational and mathematical, will be the hallmarks of the future era.

\noindent
{\bf Acknowledgements:} I wish to thank D. Baker, M. Garzelli, R. Hillerbrand, J. Kumar and J. Wacker for helpful discussions and feedback on this topic. I also wish to thank the organizers of the Wuppertal conference for a most enjoyable and stimulating event.


\begin{thebibliography}{99}

\bibitem{Harman:1965}
Gilbert H. Harman, ``Inference to the Best Explanation," The Philosophical Review 74 (1965) 88.

\bibitem{Lipton:1991}
Peter Lipton, {\it Inference to Best Explanation}, Oxford University Press, 1991.

\bibitem{Clayton:1997}
Philip Clayton, ``Inference to the Best Explanation," Zygon 32 (1997) 377.

\bibitem{Thagard:1978}
Paul R. Thagard, ``The Best Explanation: Criteria for Theory Choice", The Journal of Philosophy, 75 (1978) 76.

\bibitem{Lehrer:1974}
Keith Lehrer, {\it Knowledge}, Oxford Clarendon Press, 1974.

\bibitem{Newton-Smith:1999}
W.H. Newton-Smith, {\it The Rationality of Science}, Routledge London, 1981.

\bibitem{Kane:1996}
G. Kane, {\it The Particle Garden}, Basic Books, 1996.  

\bibitem{Griffiths:2008}
D. Griffiths, {\it Introduction to Elementary Particles}, Wiley, 2nd ed.\ (2008).

\bibitem{Richter:2006}
B. Richter, ``Theory in particle physics: theological speculation versus practical knowledge," Physics Today, October 2006.

\bibitem{Gell-Mann:1994}
M. Gell-Mann, {\it The Quark and the Jaguar}, Little, Brown and Company, 1994.

\bibitem{Gavin}
S. Gavin, J. Conn, S.P. Karrer, ``The Age of the Sun: Kelvin vs.\ Darwin," Wayne State University, 2008.

%\cite{Intriligator:2008fe}
\bibitem{Intriligator:2008fe}
  K.~Intriligator, D.~Shih, M.~Sudano,
  ``Surveying Pseudomoduli: The Good, the Bad and the Incalculable,''
  JHEP {\bf 0903}, 106 (2009).
  [arXiv:0809.3981 [hep-th]].
  
  %\cite{562147}
\bibitem{562147} 
  G.~'t Hooft,
  %``Can there be physics without experiments? Challenges and pitfalls,''
  Int.\ J.\ Mod.\ Phys.\ A\ {\bf 16}, 2895  (2001).
  %%CITATION = IMPAE,A16,2895;%%

%\cite{81239}
\bibitem{Wilson:81239} 
  K.~G.~Wilson and J.~B.~Kogut,
  %``The Renormalization group and the epsilon expansion,''
  Phys.\ Rept.\ \ {\bf 12}, 75  (1974).
  %%CITATION = PRPLC,12,75;%%

%\cite{hep-th/0602178}
\bibitem{hep-th/0602178} 
  A.~Adams, N.~Arkani-Hamed, S.~Dubovsky, A.~Nicolis and R.~Rattazzi,
  %``Causality, analyticity and an IR obstruction to UV completion,''
  JHEP\ {\bf 0610}, 014  (2006)
  [hep-th/0602178].
  %%CITATION = JHEPA,0610,014;%%
  
    %\cite{:1992ed}
\bibitem{Polchinski:1992ed}
  J.~Polchinski,
  ``Effective Field Theory And The Fermi Surface,''
  arXiv:hep-th/9210046.
  %%CITATION = HEP-TH/9210046;%%
    
\bibitem{Cohen:1993}
A.G. Cohen, ``Selected topics in effective field theories for particle physics," {\it Proceedings of the Theoretical Advanced Study Institute (TASI 1993)}, Boulder, Colorado, 1993.

\bibitem{Rothstein:2003}
I.Z. Rothstein, ``TASI lectures on effective field theories," {\it Proceedings of the Theoretical Advanced Study Institute (TASI 2003)}, Boulder, Colorado, 2003.

%\cite{Weinberg:2009bg}
\bibitem{Weinberg:2009bg}
  S.~Weinberg,
  ``Effective Field Theory, Past and Future,''
  arXiv:0908.1964 [hep-th].
  %%CITATION = ARXIV:0908.1964;%%
  
  
 \bibitem{HiggsDiscovery}
 %\cite{:2012gk}
%\bibitem{:2012gk} 
  G.~Aad {\it et al.}  [ATLAS Collaboration],
  %``Observation of a new particle in the search for the Standard Model Higgs boson with the ATLAS detector at the LHC,''
  Phys.\ Lett.\ B {\bf 716}, 1 (2012)
  [arXiv:1207.7214 [hep-ex]].
  %%CITATION = ARXIV:1207.7214;%%
%\cite{:2012gu}
%\bibitem{:2012gu} 
  S.~Chatrchyan {\it et al.}  [CMS Collaboration],
  %``Observation of a new boson at a mass of 125 GeV with the CMS experiment at the LHC,''
  Phys.\ Lett.\ B {\bf 716}, 30 (2012)
  [arXiv:1207.7235 [hep-ex]].
  %%CITATION = ARXIV:1207.7235;%%

  %\cite{arXiv:0908.2502}
\bibitem{WellsPapers} 
  M.~Olechowski {\it et al.},
  %``Reheating Temperature and Gauge Mediation Models of Supersymmetry Breaking,''
  JHEP\ {\bf 0912}, 026  (2009)
  arXiv:0908.2502.
  %%CITATION = JHEPA,0912,026;%%
  %\cite{arXiv:1009.3801}
%\bibitem{arXiv:1009.3801} 
  L.~Covi {\it et al.},
  %``Supersymmetric mass spectra for gravitino dark matter with a high reheating temperature,''
  JHEP\ {\bf 1101}, 033  (2011)
  arXiv:1009.3801.
  %%CITATION = JHEPA,1101,033;%%

%%%%%%%%%%%%%%%%%%%%%%%%%%%%%%%%%%%%%%% 

\end{thebibliography}
\end{document}